\documentclass[12pt,letter]{article}
\usepackage{scicite,times,graphicx,soul,amssymb,amsmath}
\usepackage[usenames,dvipsnames]{color}

\newenvironment{sciabstract}{%
\begin{quote} \bf}
{\end{quote}}

\newcounter{lastnote}

\title{\vspace{-0.5in} Glacial cycles drive variations in the production of oceanic crust}

\author{John W.~Crowley$^{1,2,\dag}$, Richard F.~Katz$^{1,\ast}$, Peter
  Huybers$^2$,\\ Charles H. Langmuir$^2$ \& Sung-Hyun Park$^{3,4}$\\
  \normalsize{$^{1}$Dept.~of Earth Sciences, University of Oxford, Oxford, UK}\\
  \normalsize{$^{2}$Dept.~of Earth and Planetary Sciences, Harvard University, Cambridge, USA}\\ 
  \normalsize{$^{3}$Polar Earth System Sciences, Korea Polar Research Institute, Incheon, Republic of Korea}\\ 
  \normalsize{$^{4}$Polar Climate Research, Korea Polar Research Institute, Incheon, Republic of Korea}\\
  \normalsize{$^\ast$To whom correspondence should be addressed; E-mail: Richard.Katz@earth.ox.ac.uk}\\
  \normalsize{$^\dag$Now at Engineering Seismology Group Canada, Kingston, Canada.}
}

\date{}

\begin{document} 

\baselineskip24pt

\maketitle 

\begin{sciabstract}
  Glacial cycles redistribute water between oceans and continents
  causing pressure changes in the upper mantle, with consequences for
  melting of Earth's interior. Using Plio-Pleistocene sea-level
  variations as a forcing function, theoretical models of mid-ocean
  ridge dynamics that include melt transport predict temporal
  variations in crustal thickness of hundreds of meters. New
  bathymetry from the Australian-Antarctic ridge shows significant
  spectral energy near the Milankovitch periods of 23, 41, and 100~ky,
  consistent with model predictions.  These results suggest that
  abyssal hills, one of the most common bathymetric features on Earth,
  record the magmatic response to changes in sea level. The models and
  data support a link between glacial cycles at the surface and mantle
  melting at depth, recorded in the bathymetric fabric of the sea
  floor.
\end{sciabstract}


The bathymetry of the sea floor has strikingly regular variations
around intermediate and fast-spreading ocean ridges.  Parallel to the
ridge are long, linear features with quasi-regular spacing called
abyssal hills \cite{menard67}. High resolution mapping of the sea
floor over the past few decades [e.g., refs.~\cite{scheirer96,
  macdonald96, goff97}] has shown that these hills are among the most
common topographic features of the planet, populating the sea floor
over $\sim$50,000~km of ridge length. Hypothesized models for these
features include extensional faulting parallel to the ridge
\cite{macdonald96}, variations in the magmatic budget of ridge
volcanoes \cite{kappel86}, and variation in mantle melting under
ridges due to sea-level change associated with glacial cycles
\cite{huybers09a}.  This latter model stems from the fact that
glacial-interglacial variations transfer about $5\times10^{19}$~kg of
water between the oceans and the continents.  This mass redistribution
translates to sea-level variations of $\sim$100~m and modifies the
lithostatic pressure beneath the entire ocean. Because mantle melting
beneath ridges is driven by depressurization, ocean ridge volcanism
should respond to sea-level changes, potentially leading to changes in
the thickness and elevation of ocean crust.


Plate spreading at mid-ocean ridges draws mantle flow upward beneath
the ridge; rising parcels of mantle experience decreasing pressure and
hence decreasing melting point, causing partial melting.  Mantle
upwelling rates are about 3~cm/yr on average, while sea level change
during the last deglaciation was at a mean rate of 1~cm/yr over 10,000
years.  Because water has one third the density of rock, sea-level
changes would modify the depressurization rate associated with
upwelling by $\pm10$\%, with corresponding effects on the rate of melt
production.  Mantle upwelling rate scales with the mid-ocean ridge
spreading rate, but the rate of sea level change over the global
mid-ocean ridge system is roughly uniform. On this basis, previous
workers inferred that the relative effect of sea level change should
scale inversely with spreading rate, reaching a maximum at the slowest
rates\cite{huybers09a}.  An elaboration of this model with
parameterized melt transport gave a similar scaling \cite{lund11}.

To test these qualitative inferences, we investigated the crustal
response to sea-level change using a model that computes mantle flow,
thermal structure, melting, and pathways of melt transport.  The model
is based on canonical statements of conservation of mass, momentum,
and energy for partially molten mantle\cite{mckenzie84, katz08b} and
has previously been used to simulate mid-ocean ridge dynamics with
homogeneous\cite{katz10b} and heterogeneous\cite{katz12} mantle
composition. It predicts time scales of melt transport that are
consistent with those estimated from $^{230}$Th disequilibium in young
lavas \cite{stracke06}. In the present work, the model is used to
predict crustal thickness time-series arising from changes in sea
level (Fig.~1 and \cite{sup-mats})

A suite of nine model runs for three permeability scales and three
spreading rates was driven over a 5~My period using a Plio-Pleistocene
sea-level reconstruction\cite{siddall10}. Crustal curves from
simulations with larger permeability and faster spreading rate contain
relatively more high-frequency content than lower permeability and
slower spreading rate runs (Fig.~1).  Our model results 
contradict the previous scaling arguments \cite{huybers09a,lund11} in
not showing a simple decrease in the sea level effect on ridge
magmatism with increasing spreading rate.

To better understand these numerical results, we carried out an
analysis of leading-order processes using a reduced complexity model.
This model provides a solution for crustal thickness response to
changes in sea level, approximating the results of the full numerical
model, but with greater transparency.
Assuming that all melt produced by sea level change is focused to the
ridge axis, we obtain a magmatic flux in units of kg/year per meter
along the ridge of
\begin{equation}
  M_{SL}(t) = \int_{z_m}^0 x_l(z)\:\frac{\rho_w}{\rho_m}\Pi
  \:\dot{S}\left(t-\tau(z)\right) \textrm{d} z, 
  \label{eq:mass_SL}
\end{equation}
where $\rho_w/\rho_m$ is the density ratio of sea water to mantle
rock, $\Pi$ is the adiabatic productivity of upwelling mantle (in kg
of melt per m$^3$ of mantle per meter of upwelling), $x_l(z)$ is the
half-width of the partially molten region beneath the mid-ocean ridge
at a depth $z$, and $z_m$ is the maximum depth of silicate melting
beneath the ridge. Most importantly, $\dot{S}\left(t-\tau(z)\right)$
is the rate of sea-level change $\tau$ years before time $t$
\cite{sup-mats}. 

This formulation reveals why our numerical model results (Fig.~1)
contradict earlier work \cite{huybers09a, lund11}.  Whereas earlier
work noted that variations in crustal thickness are inversely
proportional to spreading rate, $C_{SL}=M_{SL}/(U_0\rho_c)$, our model
shows that mass flux is proportional to the width of the
partially molten region beneath the ridge.  This width can be
expressed as $x_l(z)=U_0R(z)/(4\kappa)$, where $U_0$ is the
half-spreading rate, $\kappa$ is the thermal diffusivity, and $R(z)$
accounts for depth dependent influences on melting that are
independent of spreading rate (Fig.~S1).  The competing influences
associated with the volume of mantle from which melt is extracted and
the rate at which new crust is formed means that sensitivity to sea
level variation does not simply decrease with increasing spreading
rate.

Instead, the magnitude of the crustal response depends upon the
timescale of sea level forcing relative to the time required to
deliver melt from depth to the surface.  Melt
delivery times $\tau$ are computed in the reduced model using a
one-dimensional melt column formulation, and decrease with higher
pemeability and faster spreading rate \cite{hewitt10, sup-mats}.
The same response occurs in the numerical
model; in both cases, $\tau$ decreases with increasing spreading rate
because the background melting rate, dynamic melt fraction, and
permeability of the melting region all increase.

To quantify crustal response as a function of timescale, we use the
amplitude ratio of crustal to sea-level variation, called admittance,
computed at discrete frequencies by applying sinusoidal forcing.
Admittance curves for both the numerical (Fig.~2a) and reduced
(Fig.~2b) models show a distinct maximum that shifts toward higher
frequencies and larger magnitudes with shorter $\tau$.  When the
period of sea-level forcing is short relative to the characteristic
transport time $\tau_m = \tau(z_m)$, additional melt produced at depth
(falling sea-level phase) does not have time to reach the surface
before a negative perturbation to melt production occurs (rising
sea-level phase); positive and negative perturbations cancel and
crustal variation is small.  When forcing periods are long relative to
$\tau_m$, melt perturbations reach the surface but are again small
because melt production scales with the rate-of-change of sea level.
Forcing periods near $\tau_m$ give maximum admittance because of a
combination of large perturbation of melting rates and sufficient time
to reach the surface (Fig.~2c).  These results suggest that ridges are
tuned according to melt-transport rates to respond most strongly to
certain frequencies of sea-level variability.

The correspondence of the results from the numerical and reduced
models provides a sound basis for investigating the potential effects
of sea level change on sea floor bathymetry. Variations in melt
production lead to variations in crustal thickness and, through
isostatic compensation, such thickness variations should produce
changes in bathymetry identifiable in high-resolution surveys.  The
prominent spectral peaks of late Pleistocene sea level variation at
the approximately 1/100~ky$^{-1}$ ice age, 1/41~ky$^{-1}$ obliquity,
and 1/23~ky$^{-1}$ precession frequencies \cite{hays76} therefore
translate into a prediction for a bathymetric response that depends on
permeability and spreading rate.

Our model results suggest that the best chance to detect a sea level
response between 1/100~ky$^{-1}$ to 1/20~ky$^{-1}$ frequencies is at
intermediate spreading ridges.  Slow spreading ridges show little
precession response, an obliquity response that is sensitive to
uncertainties in permeability, and the effects of intense normal
faulting.  Such faulting causes rift valleys with larger relief than
expected from sea-level induced melting variations.  The sea-level
signal should be less polluted by tectonic effects at fast-spreading
ridges, but may have peak admittances at frequencies higher than
1/20~ky$^{-1}$ that would obscure the responses at predicted
frequencies.  For example, the numerical simulation with the fastest
spreading and highest pemeability has peak spectral energy at
frequencies above precession (Fig.~1b).

At intermediate half-spreading rates of 3~cm/yr, 40-ky periods lead to
predicted bathymetric variations with a wavelength of 1200~m on each
side of the ridge.  Such fine-scale variations can be obscured in
global topographic databases that grid data from multiple cruises and
may have offsets in navigation or depth. To investigate the model
predictions, a modern data set with uniform navigation and data
reduction from a single survey is preferred.  Such data is available
for two areas of the Australian-Antarctica ridge that were surveyed by
the icebreaker Araon of the Korean Polar Research Institute in 2011
and 2013 (Fig.~3).

Analysis is undertaken by identifying a region whose abyssal hill
variability is relatively undisturbed by localized anomalies,
averaging off-axis variability into a single bathymetric line, and
converting off-axis distance into an estimate of elapsed time using a
plate motion solution \cite{demets10}.  Spectral analysis of the
associated bathymetry time-series is performed using the multitaper
procedure \cite{percival93} and shows spectral peaks that are
significant at an approximate 95\% confidence level near the predicted
ice age, obliquity, and precession frequencies (Fig.~3).  Although
absolute ages are uncertain because we lack seafloor magnetic reversal
data, spectral analysis only requires constraining the relative
passage of time.  The two-sigma uncertainties associated with relative
Australian-Antarctic plate motion are $\pm$4\% \cite{demets10},
implying, for example, that the 1/41~ky$^{-1}$ obliquity signal
resides in a band from 1/39~ky$^{-1}$ to 1/43~ky$^{-1}$, a width that
is smaller than our spectral resolution.

Another check on model--data consistency is to compare magnitudes of
variability. Surface bathymetry will be roughly 6/23rds of the total
variation in crustal thickness due to the relative density differences
of crust-water and crust-mantle, assuming conditions of crustal
isostasy. The closest match between simulation results and
observations, in terms of the distribution of spectral energy, is
achieved by specifying a permeability at 1\% porosity of
$K_0=10^{-13}$~m$^2$ (Fig.~3). The standard
deviation of the simulated bathymetry is 36~m, after multiplying
crustal thickness by 6/23 and filtering \cite{sup-mats}.  To minimize
the contribution from non-sea-level induced variations in the observed
bathymetry, it is useful to filter frequencies outside of those
between 1/150$^{-1}$ and 1/10~ky$^{-1}$.  The standard deviation of
the filtered observations is 44~m, where the slightly larger value is
consistent with changes in sea level being an important but not
exclusive driver of changes in crustal thickness.

Analysis of bathymetry in another area of the Australian-Antarctic
ridge 400~km to the southeast (Fig.~S2) shows a significant spectral
peak at the obliquity frequency and indication of a peak near
1/100~ky$^{-1}$, but no peak near the precession frequencies.
Predicted and observed bathymetry is also similar with standard
deviations of 33~m and 34~m, respectively, after accounting for
fractional surface expression and filtering.  Absence of a precession
peak may result from spectral estimates being more sensitive to
elapsed time errors at higher frequencies \cite{huybers2004depth},
where such errors may be introduced through extensional faulting or
asymmetric spreading.  Detection could also be obscured by the
previously noted influence of faulting\cite{macdonald96, buck05} as
well as off-axis volcanism or sediment infilling of abyssal
troughs. Detection of significant spectral peaks at predicted
frequencies at two locations of the Australian-Antarctic ridge
nonetheless constitutes strong evidence for modulation of crustal
production by variations in sea level.


Our numerical and analytical results show a complex relationship
between spreading rate and amplitudes of crustal thickness variations
associated with changes in sea level. Perturbations to the background
melt production and delivery depend on the frequency content of the
sea-level signal, as a result of the dynamics of magma
transport. Reference mantle permeability and ridge spreading rate are
key controls on this frequency dependence.  This result could be
useful: the spreading rate can be accurately determined for a ridge,
but parameters associated with magma dynamics are far less certain,
such as the amplitude and
scaling of permeability.  Uncertainty associated with spectral
estimates of bathymetry and sea level estimates need to be better
characterized, but together these may
provide a constrain on the admittance and, thereby, dynamical
parameters of a ridge.

Although results from the high-resolution bathymetry are promising,
much remains to be done to further test the hypothesis advanced
here. Crustal thickness is not an instantaneous response to melt
delivery from the mantle, but also reflects crustal processes that may
introduce temporal and spatial averaging.  Where long-lived magma
chambers are present, for example, there may also be a crustal
time-averaging that depends on spreading rate.  In addition, faulting
at all spreading rates is an observed and important phenomenon and
sea-floor bathymetry reflects the combined effects of magma output and
crustal faulting \cite{macdonald96, buck05}. Deconvolving the relative
roles of such processes will be important.  High-resolution surveys in
targeted regions will provide the opportunity for a more complete and
rigorous analysis than is presently possible.

\bibliography{manuscript}

\begin{thebibliography}{10}

\bibitem{menard67}
H.~Menard, J.~Mammerickz, {Abyssal hills, magnetic anomalies and the East
  Pacific Rise}, {\it Earth Plan.\ Sci.\ Lett.\/} {\bf 2}, 465 (1967).

\bibitem{scheirer96}
D.~Scheirer, {\it et~al.\/}, {A Map Series of the Southern East Pacific Rise
  and Its Flanks, 15$^\circ$S to 19$^\circ$S}, {\it Marine Geophysical
  Researches\/} {\bf 18}, 1 (1996).

\bibitem{macdonald96}
K.~Macdonald, P.~Fox, R.~Alexander, R.~Pockalny, P.~Gente, {Volcanic Grothy
  faults and the origin of Pacific abyssal hills}, {\it Nature\/} {\bf 380},
  125 (1996).

\bibitem{goff97}
J.~Goff, Y.~Ma, A.~Shah, J.~Cochran, J.~Sempere, {Stochastic analysis of
  seafloor morphology on the flank of the Southeast Indian Ridge: The influence
  of ridge mophology on the formation of abyssal hills}, {\it J.\ Geophys.\
  Res.\/} {\bf 102}, 15521 (1997).

\bibitem{kappel86}
E.~Kappel, W.~Ryan, Volcanic episodicity and a nonsteady state rift-valley
  along northeast {Pacific} spreading centers: Evidence from {Sea MARC-I}, {\it
  J.\ Geophys.\ Res.\/} {\bf 91}, 13925 (1986).

\bibitem{huybers09a}
P.~Huybers, C.~Langmuir, Feedback between deglaciation and volcanic emissions
  of {CO}$_2$, {\it Earth Plan.\ Sci.\ Lett.\/}  (2009).

\bibitem{lund11}
D.~C. Lund, P.~D. Asimow, {Does sea level influence mid-ocean ridge magmatism
  on Milankovitch timescales?}, {\it Geochem.\ Geophys.\ Geosys.\/} {\bf 12},
  Q12009 (2011).

\bibitem{mckenzie84}
D.~Mc{K}enzie, The generation and compaction of partially molten rock, {\it J.\
  Petrol.\/} {\bf 25} (1984).

\bibitem{katz08b}
R.~Katz, Magma dynamics with the enthalpy method: Benchmark solutions and
  magmatic focusing at mid-ocean ridges, {\it J. Petrology\/}  (2008).

\bibitem{katz10b}
R.~Katz, Porosity-driven convection and asymmetry beneath mid-ocean ridges,
  {\it Geochem.\ Geophys.\ Geosys.\/} {\bf 10} (2010).

\bibitem{katz12}
R.~Katz, S.~Weatherley, Consequences of mantle heterogeneity for melt
  extraction at mid-ocean ridges, {\it Earth\ Planet.\ Sci.\ Lett.\/} {\bf
  335-336}, 226 (2012).

\bibitem{stracke06}
A.~Stracke, B.~Bourdon, D.~{McK}enzie, Melt extraction in the {Earth's} mantle:
  {C}onstraints from {U-Th-Pa-Ra} studies in oceanic basalts, {\it Earth Plan.\
  Sci.\ Lett.\/} {\bf 244}, 97 (2006).

\bibitem{sup-mats}
Materials and methods are available as supplementary materials on
  \textit{Science} online.

\bibitem{siddall10}
M.~Siddall, B.~Hoenisch, C.~Waelbroeck, P.~Huybers, Changes in deep {P}acific
  temperature during the mid-{P}leistocene transition and {Q}uaternary, {\it
  Quaternary Sci. Rev.\/}  (2010).

\bibitem{hewitt10}
I.~Hewitt, {Modelling melting rates in upwelling mantle}, {\it Earth Plan.\
  Sci.\ Lett.\/} {\bf 300}, 264 (2010).

\bibitem{hays76}
J.~Hays, J.~Imbrie, N.~Shackleton, {Variations in Earth's Orbit - Pacemaker of
  ice ages}, {\it Science\/} {\bf 194}, 1121 (1976).

\bibitem{demets10}
C.~{DeMets}, R.~G. Gordon, D.~F. Argus, {Geologically current plate motions},
  {\it Geophys.\ J.\ Int.\/} {\bf 181}, 1 (2010).

\bibitem{percival93}
D.~Percival, A.~Walden, {\it {Spectral Analysis for Physical Applications:
  Multitaper and Conventional Univariate Techniques}\/} (Cambridge University
  Press, 1993).

\bibitem{huybers2004depth}
P.~Huybers, C.~Wunsch, A depth-derived pleistocene age model: Uncertainty
  estimates, sedimentation variability, and nonlinear climate change, {\it
  Paleoceanography\/} {\bf 19} (2004).

\bibitem{buck05}
W.~Buck, L.~Lavier, A.~Poliakov, Modes of faulting at mid-ocean ridges, {\it
  Nature\/} {\bf 434}, 719 (2005).

\bibitem{rudge11}
J.~F. Rudge, D.~Bercovici, M.~Spiegelman, {Disequilibrium melting of a two
  phase multicomponent mantle}, {\it Geophys.\ J.\ Int.\/} {\bf 184}, 699
  (2011).

\bibitem{petsc-homepage}
S.~Balay, {\it et~al.\/}, http://www.mcs.anl.gov/petsc (2001).

\bibitem{katz07}
R.~Katz, M.~Knepley, B.~Smith, M.~Spiegelman, E.~Coon, Numerical simulation of
  geodynamic processes with the {P}ortable {E}xtensible {T}oolkit for
  {S}cientific {C}omputation, {\it Phys.\ Earth Planet.\ In.\/} {\bf 163}, 52
  (2007).

\bibitem{tabor89}
M.~Tabor, {\it Chaos and Integrability in Nonlinear Dynamics: An
  Introduction\/} (Wiley, 1989).

\bibitem{batchelor67}
G.~Batchelor, {\it An Introduction to Fluid Mechanics\/} (Cambridge University
  Press, 1967).

\bibitem{turcotte02}
D.~Turcotte, G.~Schubert, {\it Geodynamics\/} (Cambridge University Press,
  2002).

\bibitem{ribe85b}
N.~Ribe, The deformation and compaction of partial molten zones, {\it Geophys.\
  J.\ R.\ Astr.\ Soc.\/} {\bf 83} (1985).

\bibitem{hewitt09}
I.~Hewitt, A.~Fowler, Melt channelization in ascending mantle, {\it J.\
  Geophys.\ Res.\/} {\bf 114}, B06210 (2009).

\bibitem{sparks91}
D.~Sparks, E.~Parmentier, Melt extraction from the mantle beneath spreading
  centers, {\it Earth Plan.\ Sci.\ Lett.\/} {\bf 105} (1991).

\end{thebibliography}
\bibliographystyle{Science}


\paragraph{Acknowledgements} The research leading to these results has
received funding from the European Research Council under the European
Union's Seventh Framework Programme (FP7/2007--2013) / ERC grant
agreement number 279925 and from the U.S.~National Science Foundation
under grant 1338832.  Crowley thanks J.~Mitrovica and Katz thanks the
Leverhulme Trust for additional support.  Numerical models were run at
Oxford's Advanced Research Computing facility. Bathymetry data are
included in the Supporting Online Material.

\paragraph{List of supplementary content}
\begin{itemize}
\item References (21--29) are called out only in the supplementary
  content.
\item Bathymetry data files for KR1 and KR2.  Files are in columns of
  longitude (degrees east), latitude (degrees north), and elevation
  (meters).  Data format is plain text that has been compressed using
  gzip.
\item Additional information about the numerical and reduced models,
  bathymetric data acquisition and processing, and time-series
  analysis of bathymetric profiles.
\end{itemize}


\clearpage

\begin{figure}
  \centering 
  \includegraphics[width=16cm]{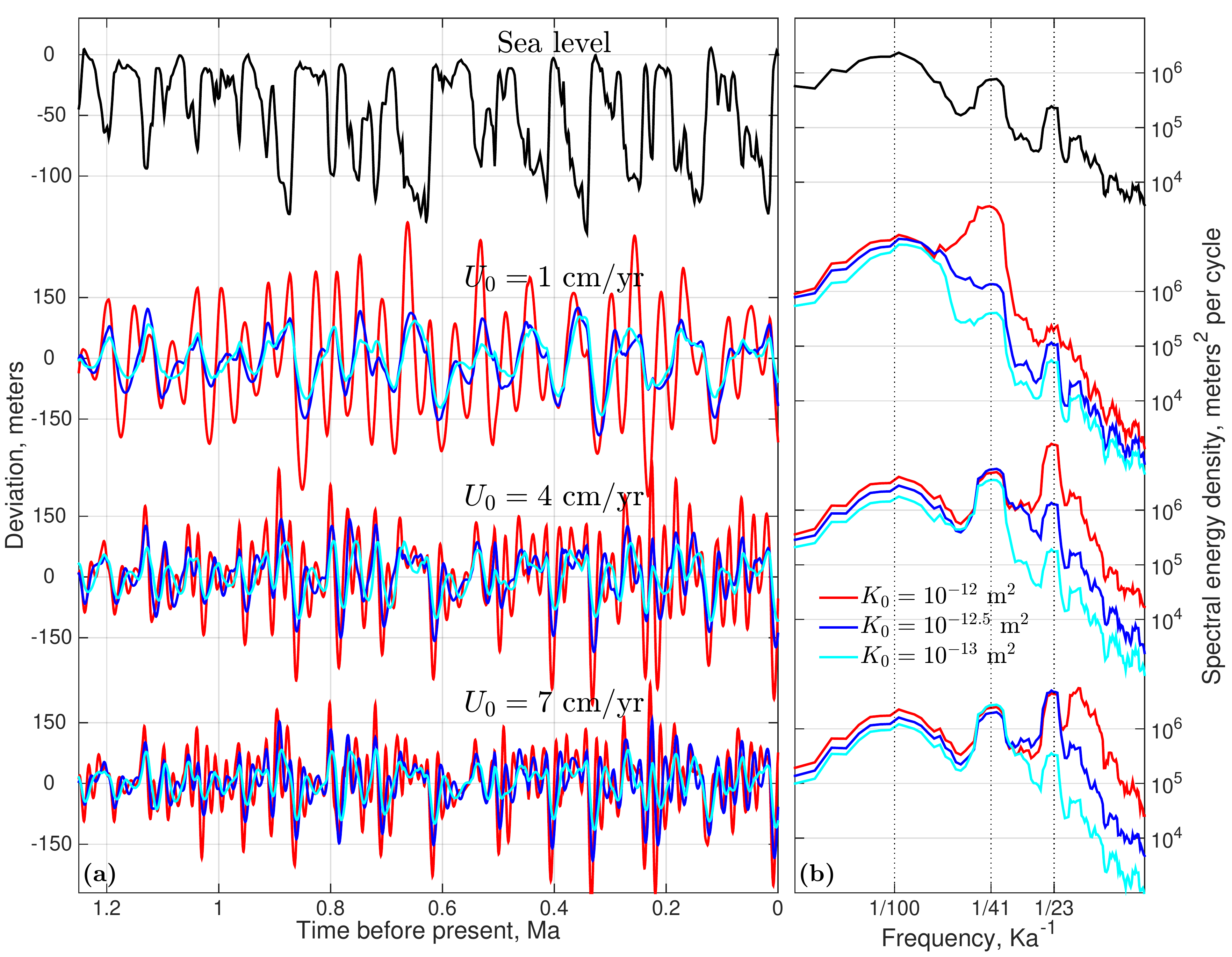}
  \caption{\textbf{Simulated bathymetric relief driven by
      Plio-Pleistocene sea-level variation.}  \textbf{(a)} Imposed
    sea-level variation (black) and predicted bathymetric relief
    (color) for the past 1.25~Ma from simulations at three
    half-spreading rates $U_0$ and three permeability levels
    $K_0$. Isostatic compensation is assumed to scale the amplitude of
    crustal thickness variation by 6/23 to give bathymetric relief.
    Permeability in the simulations is computed by applying
    $K(x,z)=K_0(\phi/\phi_0)^3$~m$^2$ to the porosity field
    $\phi(x,z)$, where $\phi_0=0.01$ is a reference porosity. Cyan,
    blue, and red lines correspond to
    $\log_{10} K_0 = -(13,\,12.5,\,12)$, respectively. \textbf{(b)}
    Power spectral density estimates for each time-series, made using
    the multitaper method with seven tapers.  Axes are logarithmic.}
  \label{fig:1}
\end{figure}

\clearpage

\begin{figure}
  \centering
  \includegraphics[width=\textwidth]{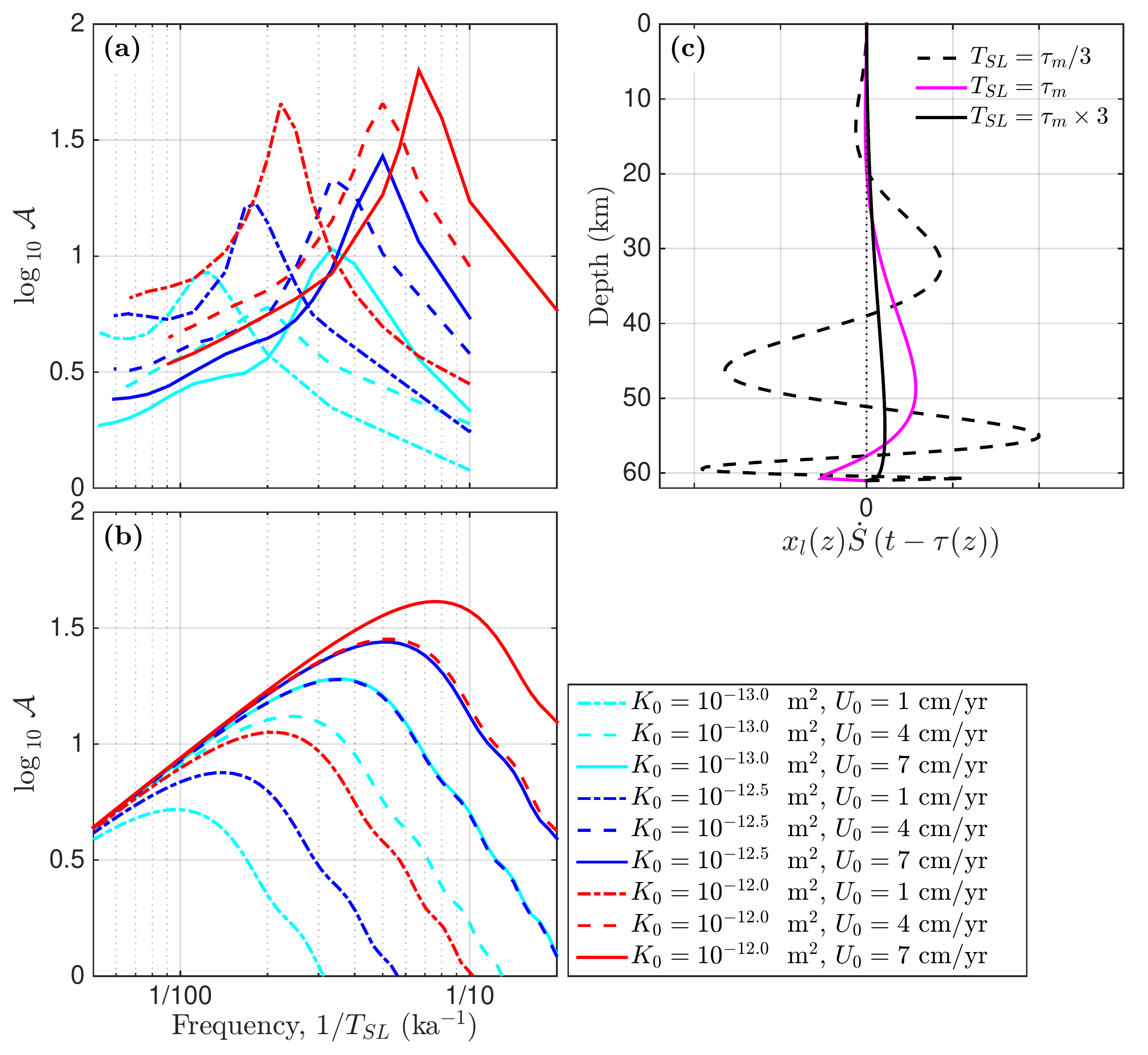}
  \caption{\textbf{Crustal thickness admittance, computed for a
      sinusoidal variation in sea level with period $T_{SL}$.}
    Admittance curves derived from \textbf{(a)} numerical simulations
    and \textbf{(b)} the reduced model \cite{sup-mats}. \textbf{(c)} A
    plot of depth $z$ versus the integrand from the reduced model of
    magma production due to sea-level variation,
    $M_{SL}(t) \propto \int_{z_m}^0 x_l(z)
    \:\dot{S}\left(t-\tau(z)\right) \textrm{d}z$
    (see also eqn.~(1) and text following).  The model is evaluated
    for for $U_0=4$~cm/yr, $K_0=10^{-13}$~m$^2$, and three values of
    sea-level oscillation period $T_{SL}$.}
  \label{fig:2}
\end{figure}

\clearpage

\begin{figure}
  \centering
  \vspace{-1in}
  \includegraphics[width=0.85\textwidth]{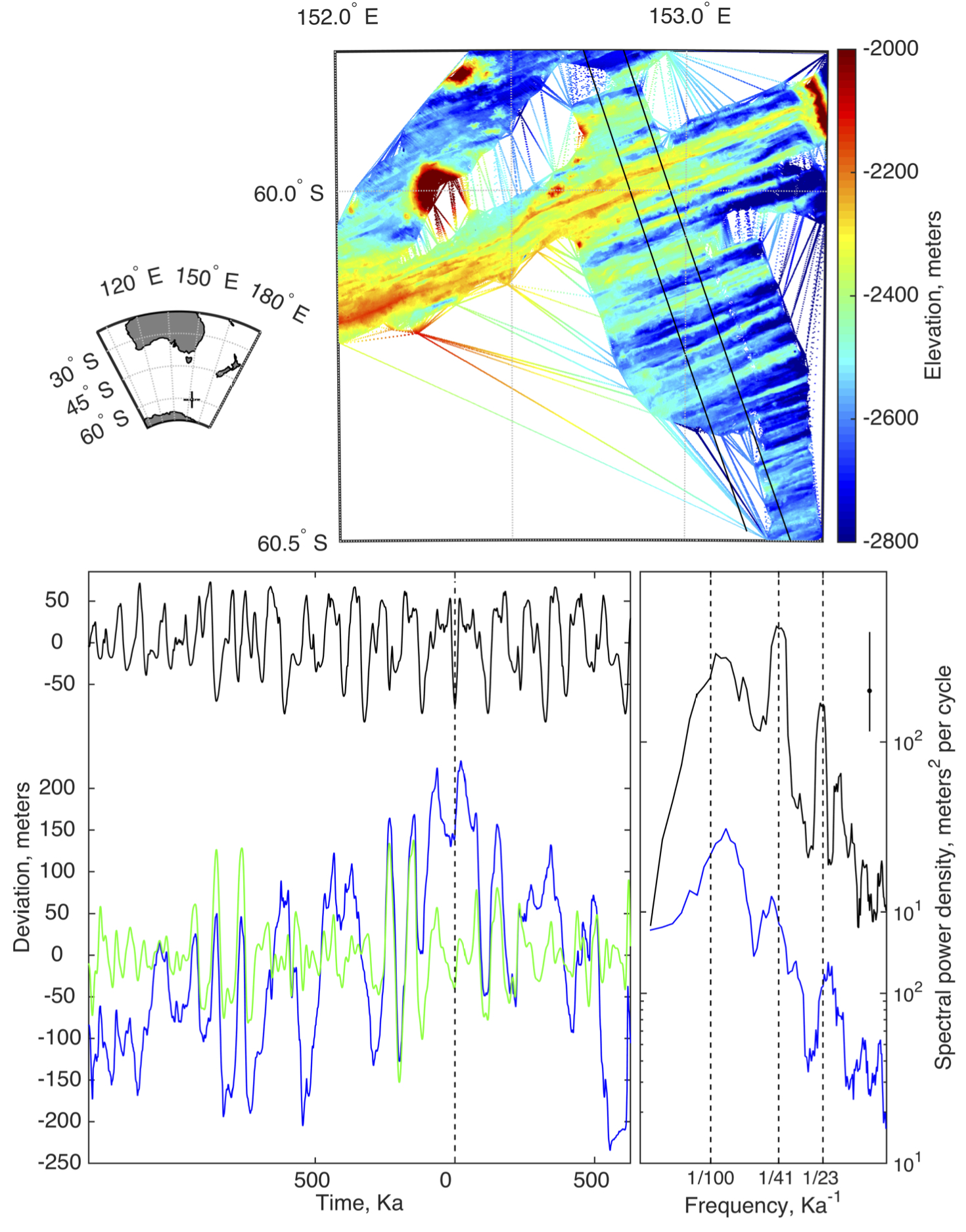}
  \caption{\small{\textbf{Bathymetry at a section of the
        Australian-Antarctic Ridge.}  A region of consistent
      bathymetry is indicated between the black lines (top right) and
      is shown in profile (bottom left, blue) after converting
      off-axis distance to an estimate of time.  Time is zero at the
      approximate ridge center.  Also shown is bathymetry after
      filtering frequencies outside of 1/150~ky$^{-1}$ and
      1/10~ky$^{-1}$ (green), and simulated bathymetry (black, for
      $U_0=3.3$~cm/yr and $K_0=10^{-13}$~m$^2$).  Spectral estimates
      (bottom right) are shown for the unfiltered bathymetry (blue)
      and model results (black), where the latter are offset upward by
      an order of magnitude for visual clarity.  Data availability is
      uneven across the ridge, and spectral estimates are for the
      longer, southern flank. Note that unlike in Fig.~1b, spectral
      estimates are prewhitened in order to improve the detectability
      of spectral peaks (see supplementary material).  Vertical dashed
      lines indicate frequencies associated with 100ky
      late-Pleistocene ice ages, obliquity, and precession.  Axes are
      logarithmic.  Statistical significance is indicated by the black
      bar at the top right of the panel: spectral peaks rising further
      than the distance between the mean background continuum
      (corresponding to the black dot) and 95th percentile (top of
      black bar) are significant.}}
  \label{fig:3}
\end{figure}

\clearpage
\appendix
\noindent{\Large Online supplementary material -- \textit{ Glacial
    cycles drive variations in the production of oceanic crust}}
\setcounter{figure}{0}
\setcounter{page}{1}
\makeatletter 
\renewcommand{\thefigure}{S\@arabic\c@figure}
\makeatother

\section{The numerical model}

A numerical model that self-consistently computes mantle flow, thermal
structure, pressure- and temperature-dependent melting, and magmatic
segregation/transport provides a context to test the hypotheses
considered here. The model is built on a theoretical framework of
conservation of mass, momentum, and energy for the magma/mantle system
\cite{mckenzie84, katz08b, rudge11}; it has been implemented in
computational simulations of mid-ocean ridges with
homogeneous\cite{katz10b} and heterogeneous\cite{katz12} mantle.  In
the present work, the same code is used with minor modifications.  The
most important change is that the lithostatic pressure at any point in
the two-dimensional domain is now augmented with a time-dependent
overburden representing sea level.  As discussed in detail below, this
is a small perturbation to the pressure, and hence the gross behavior
of the model is unaffected. Crucially, melt that is generated within
the melting regime beneath the ridge rises due to its buoyancy; some
of this melt is focused toward the ridge axis.  Melt is removed from
the domain at the ridge axis through an internal boundary representing
a sill, just below the depth where temperatures cross the solidus.
Magma percolates upward into the sill, driven by buoyancy, and is
instantaneously extracted and added to the crust.  Predicted crustal
thickness is recorded after each time-step.

The petrological model of magma genesis follows exactly from
ref.~\cite{katz10b}; melting is computed based on a linearized,
two-component phase diagram under the assumption of equilibrium
thermodynamics.  The two components represent fertile and refractory
end-member compositions, rather than particular mantle lithologies.
For simplicity and broad applicability, we only consider a chemically
homogeneous mantle source.  Details of the phase diagram, governing
equations, initialization procedure, and finite volume discretization
are previously published.  Numerical solutions to the discrete system
are obtained using the Portable, Extensible Toolkit for Scientific
Computation (PETSc, refs.~\cite{petsc-homepage, katz07}).

\begin{figure}[ht]
  \centering
  \includegraphics[width=15cm]{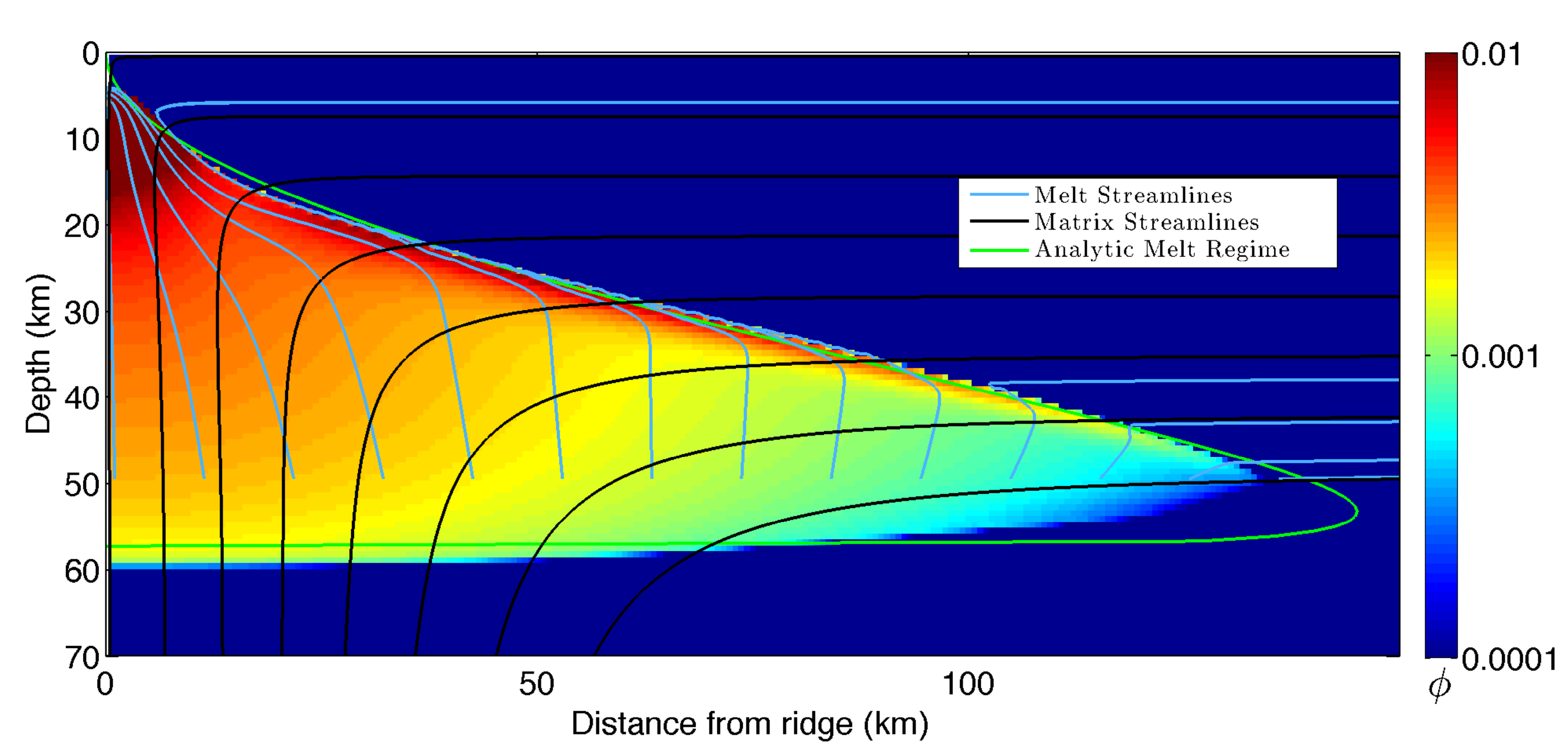}
  \caption{Numerical output for a simulation with a half-spreading
    rate of $4$ cm/yr and permeability factor $K_0=10^{-12}$ m$^2$.
    Streamlines for the magma flow (light blue) and mantle matrix
    (black) are plotted overtop of the porosity (color map).  The
    green line is the analytically determined boundary of the melt
    regime $x_l(z)$, given by equation
    (\ref{eq:app_melt_region_solidus_with_latent}).}
  \label{fig:ridgemodel}
\end{figure}

Figure~\ref{fig:ridgemodel} shows the porosity and streamlines (of
both solid mantle matrix and magma) from the numerical model of a
ridge with a half spreading rate of 4~cm/y.  A partially molten region
exists at intermediate depths between the high-pressure mantle at
depth and the low temperature lithosphere at the surface.  Mantle
material follows a corner-flow trajectory while most melt is focused
towards the ridge.  A small portion of melt, produced far from the
ridge axis, freezes into the base of the lithosphere and is
transported away laterally.

\section{The reduced model}

Analysis of numerical simulations demonstrates that the pressure
fluctuations associated with sea level produce only small
perturbations to the background, steady-state pattern of melting and
melt transport.  An approximation of this background regime, which is
associated with a constant sea level, thus forms the basis for a
reduced complexity model. Sea-level variations are then used in a
calculation of melting-rate perturbations; finally, as a leading-order
approximation, melting-rate perturbations are transported to the ridge
axis according to rates and pathways derived from the unmodified
background regime. This approach is adapted from linearized stability
analysis\cite{tabor89}, but unlike that approach, the reduced models
for the background state and perturbations are not obtained by
formally solving a linearized system of partial differential
equations.

The background state is constructed by combining aspects of the
corner-flow solution\cite{batchelor67}, the half-space cooling model
for the lithosphere\cite{turcotte02}, and a two-component melting
model\cite{ribe85b}. A one-dimensional model\cite{hewitt09} of a
column of mantle rock is used to determine the rates of melting and melt
segregation on the basis of thermodynamic and kinematic parameters of
the ridge and mantle system.  These solutions, along with a new
solution that defines the melting region, are used to solve for the
steady state production of melt by upwelling mantle flow.  Finally, an
approximate expression for the melt travel time to the ridge axis,
calculated from the column model, is used to estimate sea-level
induced variations in melt production and crustal thickness.

\subsection{Melting and melt segregation in the background state}

We first describe our approach to predicting the porosity and melt
segregation velocity that define the background state. Porosity and
melt segregation develop in response to the large scale upwelling and
melting beneath ridges.  We make several simplifications and use the
one-dimensional column model of Hewitt\cite{hewitt10}.  This model
assumes that both melt and solid flow only vertically, and that the
two phases are in local thermodynamic equilibrium.  Melting rates are
determined using conservation principles in the context of a
two-component chemical system.  For idealized, linear solidus and
liquidus relations, approximations of the melting rate, porosity, and
fluid velocity may be obtained analytically.  For a mantle solidus
given by 
\begin{equation} 
  T_S(z)=T_{S_0}-\gamma \rho_m g z+\lambda X_m,
\end{equation} 
the maximum depth of melting $z_m$ is 
\begin{equation} 
  z_m=\frac{T_{S_0}+\lambda X_m-T_m}{\gamma \rho_m
    g-\alpha g T_m/c}
  \label{app_melting_depth}
\end{equation}
where $T_{S_0}$ is the reference solidus temperature at the surface,
$\gamma$ is the Clausius-Clapeyron slope, $\lambda$ is the temperature
variation associated with changes in composition, $X_m$ and $T_m$ are
the composition and potential temperature of the deep mantle
respectively, $\rho_m$ is the mantle density, $g$ is the gravitational
acceleration, and $c$ is the specific heat.  $z$ is a coordinate
aligned with gravity that is positive upwards, such that $z=0$ defines
the surface of the solid Earth and $z_m<0$. The degree of melting, as
a function of depth, is
\begin{equation} 
  F(z) = \frac{\gamma \rho_m g c-\alpha g T_m}{L+c\lambda
    \Delta X} \left(z-z_m \right),
  \label{app_melt_fraction}
\end{equation} 
where $L$ is the latent heat of melting and $\Delta X$ is the
compositional difference between the solid and liquid phases.  The
adiabatic melt productivity of the mantle, in units of kg/m$^3$ per
meter of upwelling (i.e.~kg/m$^4$), can be calculated using
equation~(\ref{app_melt_fraction}) and is
\begin{equation} 
  \Pi=\rho_m \frac{\textrm{d}F}{\textrm{d}z}=\rho_m \frac{\gamma \rho_m g
    c-\alpha g T_m}{L+c\lambda \Delta X}.
\end{equation} 
Using this relation, the degree of melting can be expressed more
compactly as $F(z) = \Pi (z-z_m)/\rho_m$.  Assuming that compaction
stresses may be neglected throughout the column, the magmatic
upwelling rate $w_f$ is given by \cite{hewitt10}
\begin{equation} 
  w_f\approx \left(\frac{K_0 \Delta \rho g}{\phi_0^n\eta_f}
  \right)^{\frac{1}{n}} \left(\frac{\Pi
      W_m}{\rho_m}\right)^{1-\frac{1}{n}} \left( z-z_m
  \right)^{1-\frac{1}{n}},
  \label{app_melt_velocity}
\end{equation} 
where $W_m$ is the upwelling rate of the background mantle, $K_0$,
$\phi_0$, and $n$ are parameters in the permeability relation 
\begin{equation}
  \label{eq:1}
  K = K_0\left(\frac{\phi}{\phi_0}\right)^n,
\end{equation}
$\phi$ is the porosity, $\Delta \rho$ is the difference in density
between the solid and liquid phases, and $\eta_f$ is the magma
viscosity.

\subsection{Delineating the region of partial melting}

Melting and melt segregation beneath the ridge occur only in the
region where temperatures exceed the local solidus temperature. We
already know that the base of this region is $z_m$ but in
equation~(\ref{eq:mass_SL}), we require an expression for the width of
the region $x_l$ at any depth. We now develop this expression.

The potential temperature of the mantle $\tilde{T}(x,z)$ in the
vicinity of the ridge can be modelled using the half-space cooling
solution\cite{turcotte02} as
\begin{equation} 
  \tilde{T}(x,z)=T_m+\left(T_\textrm{sfc}-T_m \right)
  \textrm{erfc}\left(\frac{|z|}{2}\sqrt{\frac{U_0}{\kappa x}} \right),
  \label{app_mantle_temp_TBL}
\end{equation} 
where $\kappa$ is the thermal diffusivity, $U_0$ is the half-spreading
rate of the lithosphere at the ridge, $T_\textrm{sfc}$ is the surface
temperature, and $T_m$ is the mantle potential temperature.  The real
temperature of the mantle $T(x,z)$ is calculated by adding to
eqn.~(\ref{app_mantle_temp_TBL}) the (linearized) adiabatic
temperature gradient $\alpha g T_m z/c$.  Furthermore, in regions
where melting has occurred, the temperature is reduced by a factor of
$FL/c$, where $F$ is the degree of melt given above in equation
(\ref{app_melt_fraction}).  With these modifications, the half-space
cooling solution becomes
\begin{equation} 
  T(x,z)=T_m+\left(T_\textrm{sfc}-T_m \right)
  \textrm{erfc}\left(\frac{|z|}{2}\sqrt{\frac{U_0}{\kappa x}}
  \right)-\frac{\alpha g T_m}{c}z-\frac{\Pi L}{\rho c} (z-z_m).
  \label{app_mantle_temp_TBL_plus_adiabat_melt}
\end{equation} 
Using a mantle solidus temperature \cite{hewitt10}
\begin{equation} 
  T_S(z)=T_{S_0}-\gamma \rho g z+\lambda \chi_m,
\end{equation} 
we can solve for the curve defining the boundary of the melt region
$x=x_l(z)$ by setting $T(x_l,z)$ from equation
(\ref{app_mantle_temp_TBL_plus_adiabat_melt}) equal to $T_S(z)$.  This
gives
\begin{equation} 
  \label{eq:app_melt_region_solidus_with_latent}
  x_l(z)=\frac{U_0 z^2}{4 \kappa} \left(
    \textrm{erfc}^{-1} \left[\frac{\gamma \rho g-\alpha g T_m/c-\Pi L/\rho
        c}{T_m-T_\textrm{sfc}} \left(z-z_m\right) \right] \right)^{-2}.
\end{equation} 
This can be written compactly as
\begin{equation} 
  \label{app_melt_region_solidus_with_latent_short}
  x_l(z)=\frac{U_0}{4 \kappa} R(z),
\end{equation} 
where $R(z)$ is a function that depends on the thermodynamic
parameters of the mantle but is independent of spreading rate and
permeability:
\begin{equation} 
  R(z)=\left( \frac{z}{ \textrm{erfc}^{-1}
      \left[\frac{\gamma \rho g-\alpha g T_m/c-\Pi L/\rho c}{T_m-T_\textrm{sfc}}
        \left(z-z_m\right) \right]  }\right)^{2}.
  \label{app_R_z}
\end{equation} 
Equation (\ref{app_melt_region_solidus_with_latent_short})
demonstrates that the cross-sectional area of the melt region scales
linearly with the half-spreading rate.

\subsection{The melt transport time}
\label{sec:melt-travel-time}

We expect sea-level induced variations in melting rate to be small
compared to melting rates associated with upwelling mantle.  We
therefore assume that the rate of melt segregation, calculated using
the column model for the background decompression melting, will remain
unchanged.  This rate can then be used to calculate the transport time
for perturbations in melt production due to changing sea level.
The one-dimensional melt transport rate for a parcel of fluid is
$w_f=\textrm{d} z/\textrm{d} t$.  Rearranging, integrating, and using
equation~(\ref{app_melt_velocity}) for the melt speed gives
\begin{eqnarray} 
  \tau(z) &=& \int_z^{0} \frac{\textrm{d} z}{w_f(z)}, \nonumber \\
  &=& \int_z^{0} \left(\frac{\eta_f\phi_0^n}{K_0 \Delta \rho g}
  \right)^{\frac{1}{n}} \left(\frac{\Pi
      W_m}{\rho}\right)^{\frac{1}{n}-1} \left( z-z_m \right)^{\frac{1}{n}-1}
  \textrm{d} z, \nonumber \\ &=& n \left(\frac{\eta_f\phi_0^n}{K_0 \Delta \rho g}
  \right)^{\frac{1}{n}} \left(\frac{\Pi
      W_m}{\rho}\right)^{\frac{1}{n}-1} \left[
    (-z_m)^{\frac{1}{n}}-(z-z_m)^{\frac{1}{n}} \right], \nonumber \\ &=&
  \tau_m \left[ 1-\left( 1-\frac{z}{z_m} \right)^{\frac{1}{n}} \right].
  \label{app_tau_equation}
\end{eqnarray} 
$\tau(z)$ is the time taken for a parcel of material
that melts at a depth $z$ to reach the surface.  $\tau_m$ is the time
taken for melt at the base of the melting column ($z=z_m$) to reach
the surface and is given by
\begin{equation} 
  \tau_m=n \left(\frac{\eta_f\phi_0^n}{K_0 \Delta \rho g}
  \right)^{\frac{1}{n}} \left(\frac{\Pi
      W_m}{\rho}\right)^{\frac{1}{n}-1} (-z_m)^{\frac{1}{n}}.
  \label{app_tau_m}
\end{equation}

For simplicity we will assume that the melt travel time is independent
of lateral position $x$ and that the solution for a one-dimensional
column can be applied independently of lateral position.  This
assumption simplifies the calculation considerably. More sophisticated
transport models are possible, but at the cost of reliance on numerical
methods.  However, since we already have a full numerical solution to
the governing equations, we have sought a reduced model that is
computed analytically, captures only the leading-order physics, and is
hence easily interpretable.

This simplification excludes two main factors in the melt transport
process.  It gives no consideration for the lateral flow of melt that
would be necessary to focus off-axis melting towards the ridge.
\cite{sparks91} demonstrated that a high-porosity boundary layer forms
at the base of the lithosphere, creating a channel that rapidly
transports melt laterally towards the ridge axis.  Melt is therefore
expected to flow sub-vertically beneath the high-porosity boundary
layer; the column model represents a reasonable approximation for this
flow.  

An additional, useful relationship is the depth as a function of delay
time.  This is given by inverting equation (\ref{app_tau_equation})
for $z$ and defining $\zeta(\tau)$ as
\begin{equation} 
  \zeta(\tau) = z_m \left[1-\left( 1-
      \frac{\tau}{\tau_m} \right)^n \right].
  \label{app_z_tau_relation}
\end{equation}

\subsection{Crustal thickness variations due to sea-level change}

We now develop a calculation of the effect of changes in sea-level on
crustal thickness.  Let $S(t)$ represent the sea-level height at time
$t$ with respect to some fixed, reference sea level.  Pressure
variations associated with sea-level changes will, to leading order,
be felt equally everywhere beneath the ridge and are assumed to be
independent of spatial position in the mantle.  The perturbation to
the melting rate from sea level change is
\begin{equation} 
  \Gamma_{SL}=\frac{\rho_w}{\rho_m}\Pi\dot{S}.
  \label{app_sea_level_melt_pert}
\end{equation} 
where $\dot{S}=\textrm{d}S/\textrm{d}t$.  Sea-level induced
perturbations in melting rate will be transported to the ridge by the
background magma flow.  A perturbation at time $t$ and position
$(x,z)$ will arrive at the ridge and be recorded in the crustal
thickness at a time $t+\tau (x,z)$, where $\tau (x,z)$ is the time
taken for the melt to travel to the surface.  The total melt delivery
to the ridge from sea-level induced melting $M_{SL}$ in units of
kg/year per meter along the ridge is
\begin{equation}
  M_{SL}(t)=\int_{z_m}^{0 } \int_0^{x_l(z)} \frac{\rho_w}{\rho_m}\Pi
  \,\dot{S}\left(t-\tau(x,z)\right)\,\textrm{d}x\,\textrm{d}z ,
  \label{app_mass_SL}
\end{equation}
where $x_l(z)$ is the distance between the ridge axis and the edge of
the melting regime.  By assuming that the travel time depends only on
the depth $z$ (see sec.~\ref{sec:melt-travel-time}, above), we can
carry out the inner integral to arrive at the approximate and simpler
expression
\begin{equation} 
  M_{SL}(t)\approx \frac{\rho_w}{\rho_m}\Pi \int_{z_m}^{0} x_l(z)
  \dot{S}\left(t-\tau(z)\right) \,\textrm{d}z .
  \label{app_mass_SL_2}
\end{equation} 
To compute the crustal thickness variation associated with this mass
delivery rate we divide $M_{SL}$ by the crustal density and the ridge
half-spreading rate to give
\begin{equation} 
  C_{SL}(t)= \frac{\Pi}{U_0} \frac{\rho_w}{\rho_m \rho_c}
  \int_{z_m}^{0} x_l(z) \dot{S}\left(t-\tau(z)\right)\, \textrm{d}z .
  \label{app_SL_crustal}
\end{equation} 
Equation (\ref{app_SL_crustal}) requires the geometry of the melting
region $x_l(z)$ from
equation~(\ref{app_melt_region_solidus_with_latent_short}) and the
travel time for melt produced at depth to rise to the surface
$\tau(z)$ from equation~(\ref{app_tau_equation}). The inverse
dependence of $C_{SL}$ on $U_0$ cancels with the linear dependence of
$x_l$ on $U_0$; however, admittance in Fig.~2c still depends
on spreading rate because of the spreading-rate control on melt
travel-time $\tau$ (eqns.~(\ref{app_tau_equation}) and
(\ref{app_tau_m})).  

\subsection{A Green's function for crustal thickness perturbations}

The crustal thickness response to a instantaneous step-change in sea
level is quantified by a Green's function; convolution of this
function with an arbitrary sea-level history $\dot{S}$ provides a
convenient way to compute the predicted variation in crustal
thickness.  To calculate the Green's function for the crustal
thickness we calculate the response of the system to an instantaneous,
unit change in sea level at time $t=0$.  The sea level as a function
of time is then $S=S_0+ H(t),$ where $H(t)$ is the Heaviside function
(equal to zero for $t<0$ and one for $t \ge 0$).  The rate of change
of sea-level is then simply given by the Dirac delta-function,
\begin{equation} 
  \dot{S}=\delta (t).
  \label{app_sea_level_rate}
\end{equation} 
From equation (\ref{app_SL_crustal}) the crustal
thickness is given by
\begin{equation} 
  C_{\delta}(t)= \frac{\Pi}{U_0}
  \frac{\rho_w}{\rho_m \rho_c} \int_{z_m}^{0} x_l(z) 
  \dot{S}\left(t-\tau(z)\right) \textrm{d} z.
  \label{app_SL_crustal_2}
\end{equation}
Substituting in equations
(\ref{app_melt_region_solidus_with_latent_short}) and
(\ref{app_sea_level_rate}) for $x_l(z)$ and $\dot{S}$ and simplifying
gives
\begin{equation} 
  C_{\delta}(t)=\frac{\Pi}{4 \kappa}
  \frac{\rho_w}{\rho_m \rho_c} \int_{z_m}^{0} R(z)\, \delta(t-\tau(z))\, \textrm{d} z .
  \label{app_SL_crustal_3}
\end{equation} 
The integration variable can be switched from depth to time using
$\textrm{d}z = \zeta' \textrm{d} \tau$, where $\zeta'=\textrm{d}
\zeta/\textrm{d} \tau$ can be calculated from equation
(\ref{app_z_tau_relation}), giving
\begin{equation} 
  C_{\delta}(t)=\frac{\Pi}{4 \kappa}
  \frac{\rho_w}{\rho_m \rho_c} \int_{\tau_m}^{0} R\left(\zeta(\tau)
  \right)\, \delta(t-\tau)\,  \zeta'(\tau) \textrm{d} \tau .
  \label{app_SL_crustal_4}
\end{equation} 
Due to the presence of the delta function, the integration can be
carried out to give
\begin{equation} 
  C_{\delta}(t)=
  \begin{cases}
    -\frac{\Pi}{4 \kappa}
    \frac{\rho_w}{\rho_m \rho_c} R\left( \zeta(t) \right) \zeta'(t) &
    \text{for $0\leq t\leq \tau_m$,}\\
    0 & \text{for $t<0$ and $t>\tau_m$}.
  \end{cases}
  \label{app_SL_crustal_5}
\end{equation} 

We have assumed that the melting-rate perturbations caused by
SL-variation do not alter the background state and hence the crustal
response to an arbitrary sea-level history can be obtained by
decomposing that history into a series of scaled Dirac delta functions
and superposing the response to those
impulses. Equation~(\ref{app_SL_crustal_5}) gives the response to an
individual impulse, and therefore the solution at time $t$ for an
arbitrary sea-level forcing function is
\begin{equation} 
  C_{SL}(t)=\int_{-\infty}^t C_{\delta}
  \left(t-\tilde{t}\,\right) \dot{S}
  \left(\tilde{t}\,\right) \textrm{d} \tilde{t}.
  \label{app_C_SL}
\end{equation}
In practice, the limits of integration can be reduced to
$[t-\tau_m,\,t]$ because $C_\delta$ is non-zero only in that interval.
Equation (\ref{app_C_SL}) sums all contributions to the crustal
thickness from sea-level change from $t=-\tau_m$ to time $t$.  Then by
examination, we find that the Green's function is
\begin{equation} 
  \mathcal{G}\left(t, \tilde{t}\,\right)=C_{\delta}
  \left(t-\tilde{t}\,\right)=-\frac{\Pi}{4 \kappa} \frac{\rho_w}{\rho_m \rho_c}
  R\left( \zeta\left(t-\tilde{t}\,\right) \right) \zeta'\left(t-\tilde{t}\,\right),
  \label{app_greens_fctn}
\end{equation} 
where $\zeta(t)$ is given by eqn.~(\ref{app_z_tau_relation}),
$R(\zeta)$ is given by eqn.~(\ref{app_R_z}), and $\zeta'(t)$ is
obtained by taking the derivative of $\zeta$ with respect to $\tau$.
The crustal thickness for an arbitrary sea-level forcing is then given
by the convolution
\begin{equation} 
  C_{SL}(t)=\int_{t-\tau_m}^t \mathcal{G}\left(t, \tilde{t}\,\right) \,
  \dot{S}\left(\tilde{t}\,\right) \textrm{d} \tilde{t}.
  \label{app_greens_integral}
\end{equation}

\subsection{Comparison with the full model}

\begin{figure}[ht]
  \centering
  \includegraphics[width=8cm]{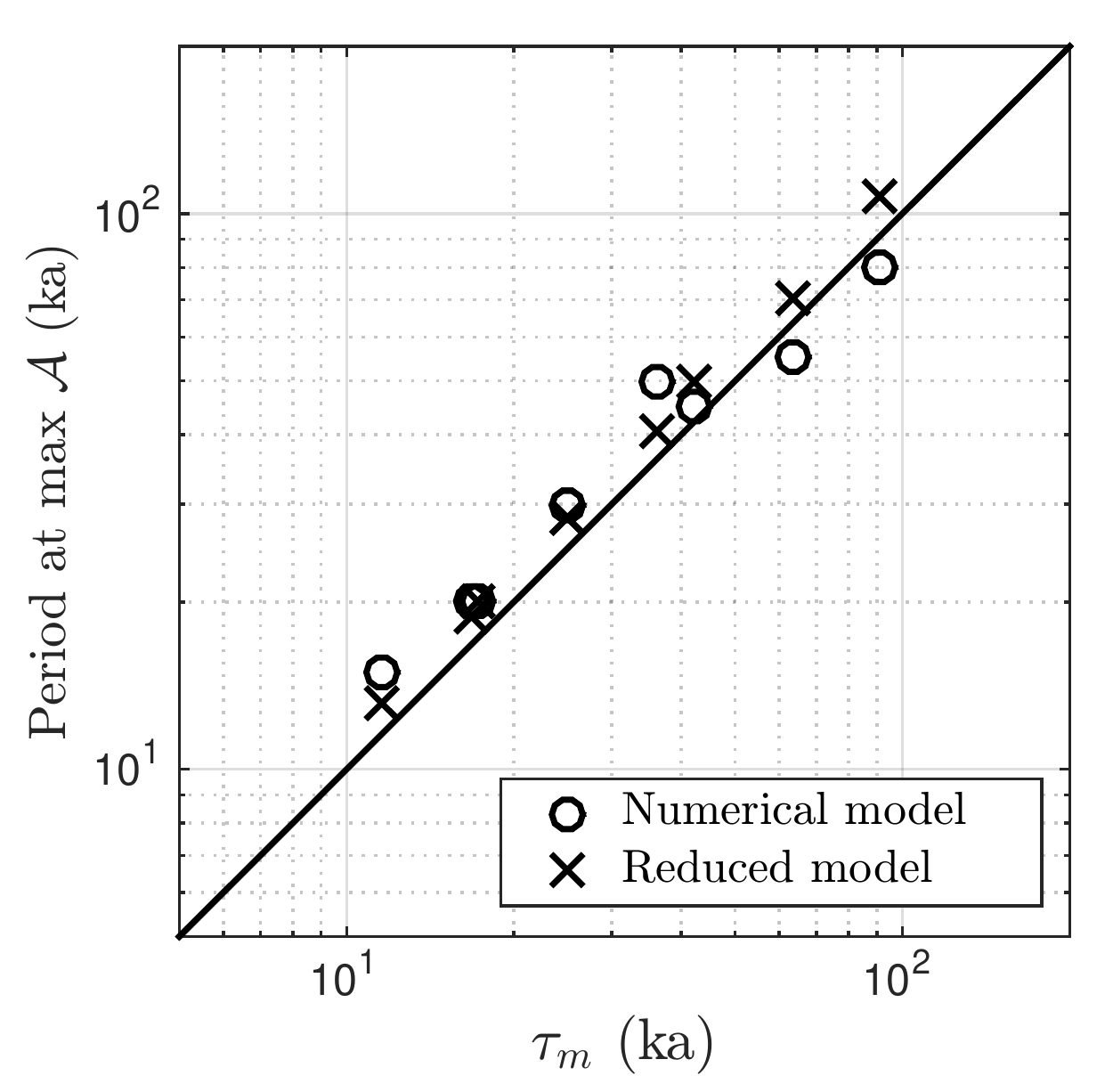}
  \caption{Comparison of the full and reduced model in terms of the
    sea-level period that gives the maximum admittance
    $\mathcal{A}$. The $x$-axis is the melt transport time computed
    using the reduced model.}
  \label{fig:full_reduced_compare}
\end{figure}

Admittance curves computed with the full, numerical model and with the
reduced model are show in Figures~\ref{fig:2}(a) and (b),
respectively.  A key characteristic of these curves is the forcing
period at which they reach peak admittance.
Figure~\ref{fig:full_reduced_compare} shows the period at peak
admittance for the full and the reduced models. The good agreement
indicates that the reduced model captures the physics that controls
the magmatic response to sea-level variation.

\section{Bathymetric data acquisition and processing}

The icebreaker Araon of the Korean Polar Research Institute is
equipped with an EM~122 multi-beam echosounder from Kongsberg for
measuring bathymetry. The device uses up to 288 simultaneous beams per
swath and operates at about 12~kHz.  Data is processed using the HIPS
\& SIPS software (version 7.0) of the CARIS company.

\section{Time-series analysis of bathymetry profiles}

We first seek to identify a sequence of bathymetry from the
Australian-Antarctic ridge that is broadly representative of the
abyssal hill structure at the ridge.  Lines traversing the ridge are
defined by an Euler pole from the MORVEL plate spreading solution
\cite{demets10} at approximately 1~km intervals.  Bathymetry
observations are spaced at approximately 50~m in this region and
values along the lines are obtained by interpolation using a Delaunay
triangulation.  Higher resolution sampling of lines has no appreciable
influence on the spectral results reported below.

Distance along lines is converted into time in the past by dividing by
spreading rate, where the half spreading rate is about 3.3~cm/yr and
account is taken of variations associated with distance from the Euler
pole.  In order to align bathymetry lines, cross-covariance is
computed between each successive pair, and the more eastward line is
shifted into a position that maximizes covariance.  For purposes of
aligning the structure of interest, it is useful to filter each line
for variations at frequencies outside of those between 1/150~ky$^{-1}$
and 1/10~ky$^{-1}$.  A zero-phase backward-forward method is used for
filtereing with a seven-point Butterworth filter.  Cross-correlations
between successive pairs of filtered and aligned bathymetry lines show
a region between 152.9$^\circ$E and 153.1$^\circ$E comprising 7 lines
whose correlations are all above 0.7.  Only two other instances
amongst 20 have comparable correlations, suggesting that this region
is especially pristine.  The aligned but unfiltered versions of these
7 lines are averaged together to form a single time series.

To estimate the center-point of the ridge in the average line, an
initial guess of the highest point is selected.  For purposes of
display, we plot bathymetry from south of the ridge on the left and
northern bathymetry to the right.  According to our model, bathymetry
variations resulting from changes in sea level should be symmetric
across the ridge, and we search within $\pm$50~ky of the selected high
point for a center about which the cross-correlation of the segments
north and south of the ridge are maximised.  The bathymetry line is
again filtered outside of frequencies between 1/150~$ky^{-1}$ and
1/10~$ky^{-1}$ in order to focus on frequencies where sea level
variability is most energetic.  Adding 22~ky to the age of the
bathymetry line maximizes cross-correlation across the ridge, giving a
value of 0.50, and has the effect of shifting the estimated ridge
center southward.


The spectrum of ridge bathymetry is estimated using the multitaper
method with seven tapers \cite{percival93}.  We focus on the segment
south of the ridge because it extends 1310~ky, as opposed to 629~ky
north of the ridge, and because the longer duration permits for
greater frequency resolution.  Spectral energy density strongly
increases toward lower frequencies, as follows from the effects of
thermal subsidence, the presence of an axial rise, and likely as a
consequence of faulting.  In order to better identify spectral peaks
amidst this red background continuum, time series are pre-whitened
prior to spectral analysis by taking the time difference.
Pre-whitening gives nearly equivalent results to multiplying
unwhitened spectral estimates by frequency squared but has the benefit
of making each tapered spectral estimate more independent.  It follows
that pre-whitening increases the estimated equivalent degrees of
freedom from an average of 12.07 to 13.99 at frequencies between 1/150
and 1/10~ky$^{-1}$, where the upper bound is 14, or twice the number
of tapered spectral estimates.

To assess statistical significance we adopt a null hypothesis of a
smoothly varying background spectral energy density that randomly
varies according to a chi-squared distribution having the estimated
nearly 14 degrees of freedom.  The approximate 95th percentile of this
null is indicated by the black bar at the upper right of Fig.~3c.
Statistically significant spectral peaks are indicated at frequencies
where the estimated spectral energy density rises above the mean
background continuum (represented by the dot) by a distance greater
than that to the 95th percentile of the null (top of the bar).  The
lower extension of the bar indicates the distance between the mean and
5th percentile.  Note that the use of logarithmic scaling in the
ordinate permits for applying this confidence interval with a constant
vertical distance because the mean and variance of the chi-square
distribution has a constant linear relationship.  Spectral peaks at
frequencies near the main Pleistocene ice age cycle (1/100~ky$^{-1}$),
obliquity (1/41~ky$^{-1}$), and precession (1/23~ky$^{-1}$) are each
indicated as being statistically significant, mirroring the spectral
peaks predicted by our model in response to sea level variations.

Bathymetry at another section of the Australian-Antarctic ridge 400~km
southeast is analysed using the same approach, where 20 lines between
158.7$^\circ$E and 159.2$^\circ$E are identified as being especially
consistent (see supplementary Fig.~S2).

\begin{figure}
  \centering
  \vspace{-1in}
  \includegraphics[width=\textwidth]{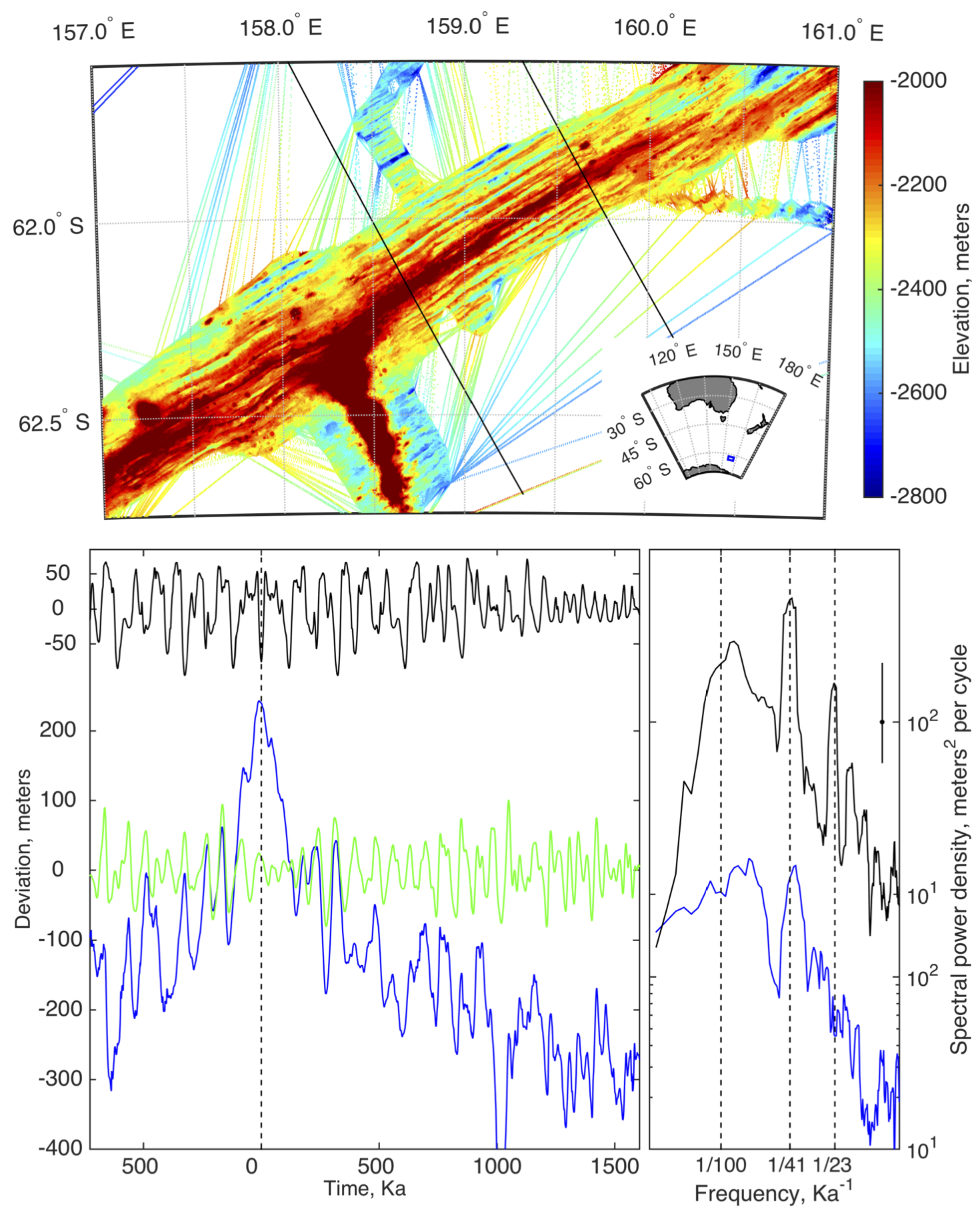}
  \caption{Similar to Fig.~3 but for a section of the
    Australian-Antarctic Ridge 400 km to the southeast.  A region of
    consistent bathymetry is indicated between the black lines (top)
    and shown in profile after converting off-axis distance to an
    estimate of time (bottom left, blue) and after filtering (green).
    Simulated bathymetry is the same as that shown in Fig.~3
    ($U_0=3.3$~cm/yr and $K_0=10^{-13}$~m$^2$).  Spectral estimates
    (bottom right) are shown for the full bathymetry (blue) and model
    results (black, offset upward by an order of magnitude).}
  \label{fig:3}
\end{figure}


\begin{table}[p]
  \begin{center}
    \begin{tabular}{c|rl|l}
      \hline     
      Parameter		&	Value	&	&	Parameter description					\\
      \hline
      $\rho$		&	$3000$	& 	kg/m$^3$        &	Reference density					\\
      $\rho_w$		&	$1000$	& 	kg/m$^3$                &	Density	of water				\\
      $\Delta \rho$	&	$500$	& 	kg/m$^3$       &  Mantle--magma density difference				\\
      $g$		&	$10$		&	m/s$^2$		&	Gravitational acceleration		\\
      $K_0$		&	$10^{-13}$	&	m$^2$           &	Reference permeability		\\
      $\phi_0$          &       $0.01$          & & Reference porosity \\
      $K$		&	$K_0 (\phi/\phi_0)^n$	&	m$^2$
      &	Permeability at porosity $\phi$		\\
      $\eta_l$		&	$1$		&	Pa$\cdot$s	&	Fluid/melt viscosity			\\
      $L$		&	$4\times 10^{5}$	&	J/kg	&	Latent heat				\\
      $c$		&	$1200$	&	J/kg$\cdot$K	&	Specific heat				\\
      $\kappa$		&	$10^{-6}$	&	m$^2$/s	&	Thermal diffusivity	\\
      $\alpha$		&	$3\times 10^{-5}$	&	1/K	&	Thermal expansion			\\
      $\gamma$          &	$60\times 10^{-9}$	&	K/Pa		&	Clapeyron slope		\\
      $\lambda$		&	$400$	&	K	&	Solidus change due to composition 			\\
      $\chi_m$		&	$0.85$	&			&	Mantle composition				\\
      $\Delta \chi$	&	$0.1$	&			&	Solid-liquid composition difference	\\
      $T_m$		&	$1648$	&	K		&	Potential temperature of upwelling mantle	\\
      $T_{s_0}$		&	$1565-\lambda \chi_m$	&	K		&	Reference solidus temperature	\\
      $W_m$		&	various	&	cm/y		&	Upwelling mantle velocity	\\
      $n$		&	$3$	&			&	Exponent in porosity relation	\\
      $\Delta S$	&	$-100$	&	m		&	Sea level change amplitude \\
      \hline
    \end{tabular}
  \end{center}
  \caption{Parameter values for calculations.}
  \label{tab:params}
\end{table}

\end{document}